# openSE: A SYSTEMS ENGINEERING FRAMEWORK PARTICULARLY SUITED TO PARTICLE ACCELERATOR STUDIES AND DEVELOPMENT PROJECTS*

P. Bonnal [†], B. Féral, K. Kershaw, B. Nicquevert, CERN, 1211 Geneva 23, Switzerland
M. Baudin, École Normale Supérieure, UPMC & CNRS UMR 7203 LBM, 75005 Paris, France
L. Lari[1], Fermilab, Batavia IL 60510, USA
J. Le Cardinal, École Centrale/SupÉlec, 92290 Châtenay-Malabry, France
[1] also at European Spallation Source, 221 00 Lund, Sweden

*Abstract*

Particle accelerator projects share many characteristics with industrial projects. However, experience has shown that best practice of industrial project management is not always well suited to particle accelerator projects. Major differences include the number and complexity of technologies involved, the importance of collaborative work, development phases that can last more than a decade, and the importance of telerobotics and remote handling to address future preventive and corrective maintenance requirements due to induced radioactivity, to cite just a few. The openSE framework it is a systems engineering and project management framework specifically designed for scientific facilities' systems and equipment studies and development projects. Best practices in project management, in systems and requirements engineering, in telerobotics and remote handling and in radiation safety management were used as sources of inspiration, together with analysis of current practices surveyed at CERN, GSI and ESS.

## INTRODUCTION

The conception and development of large-scale scientific facilities emitting ionizing radiations rely more on project management practices in use in the process industry than on systems engineering practices. This paper aims at briefly highlighting possible reasons for this present situation and to propose a way to enhance systems engineering so that the specific radiation safety requirements are considered and integrated in the approach. We have reviewed lessons learned from the management of large-scale scientific projects, and more specifically that of the Large Hadron Collider project at CERN. It is shown that project management and systems engineering practices are complementary and can beneficially be assembled in an integrated and lean managerial framework that grants the appropriate amount of focus to safety and radiation safety aspects.

## SCIENTIFIC PROJECT MANAGEMENT

*Scientific Project Specificities*

Even if some professional associations such as the US Project Management Institute [1] aim to homogenize practices, the project management corpus differs substantially depending on the professional domain it is applied to. One can typically observe different approaches for the eight following domains: construction, process industry, new product development, new service development, information and communication technologies, organization, events and human resource development. Research projects shall be considered separately as they do not completely fulfil all of the generally agreed definitions for projects in organizations. So-called scientific projects differ from research projects in the sense that they are aimed at providing means for performing research projects. In many research domains, scientific projects share many of the characteristics of industrial projects. This is the case for particle accelerator facilities that have many engineering aspects in common with process industry facilities, such as chemical plants or power stations. While "reasonable-size" projects belong typically to one of the eight domains mentioned above, large-scale scientific facilities are composed of sub-projects from all eight domains [2].

The conception, development and construction of large-scale scientific facilities rely on the appropriate use of project management practices. However, these practices are not unique; they are many and specific to certain aspects of the project. Forcing all project contributors to implement a unique project management approach must have a rationale. Sharing a common core is a prerequisite for enhancing communication and coordination among project participants, although the definition of this is not straightforward. Some suggest the implementation of PMI's Project Management Body of Knowledge [3], which aims to describe project management practices suited to all types of projects. The recently released ISO 21500 Std. [4] "can be used by any type of organization, including public, private or community organizations, and for any type of project, irrespective of complexity, size and duration". Both documents are general-purpose standards, but are not sufficiently specific to fulfil the expectations mentioned above. For that very reason, they can serve as a basis for the core framework, but cannot be the core framework. While the NASA's Systems Engineering Handbook [5], or the ESA's ECSS Standards [6], propose such a core framework for space projects, to our knowledge, nothing similar exists for large-scale particle accelerator projects, and more broadly for scientific projects.

---

* Work supported by European Commission, FP7 (G.A. no. 264336).
† email address: Pierre.Bonnal@cern.ch

*Systems Engineering*

According to the International Council on Systems Engineering (INCOSE) [7], "systems engineering (SE) is an interdisciplinary approach and means to enable the realization of successful systems. It focuses on defining customer needs and required functionalities early in the development cycle, documenting requirements, and then proceeding with design synthesis and system validation while considering the complete problem" [8]. In other words, SE can be seen as a subset of the project management corpus dedicated to the development of complex mechatronics systems.

Several academic studies confirmed that the implementation of standardized practices increases project success (see for instance Milosevic and Patanakul [9]). But other studies convey the contrary, such as that of Dvir, Raz and Shenhar [10]. To these authors: "The findings suggest that project success is insensitive to the level of implementation of management processes and procedures, which are readily supported by modern computerized tools and project management training. On the other hand, project success is positively correlated with the investment in requirements' definition and development of technical specifications". The SE corpus is somehow in line with this conclusion: it suggests that particular attention be paid to the technical side of the project.

*The ORAMS Requirements*

Out of the four concerns of ORAMS (Operability, Reliability, Availability, Maintainability and Safety), safety may impact the success of a project quite differently from the others because of the way its deliverables are assessed. While the four ORAM requirements are associated with tangible deliverables, the "safety success" of a system relies on intangible deliverables. If, over its development and operations, there is no impact on people, no accident, and no impact on the environment, then a project is considered a success. However, in terms of ORAM requirements, whose assessment is based on something happening, the "safety success" will not be seen because "nothing" tangible has happened. For this reason, stakeholders are usually keener to work out the ORAM side of the project rather than the safety side. As a result, the five ORAMS requirements should be addressed differently.

As exposed in the literature [5, 8, 11], systems engineering (SE) does not provide the means to handle this specificity. For instance, although the NASA SE Standard suggests that safety reviews be organized regularly, out of the 33 typical activities of the Concept and Technology Development Phase of a space project [5] (p.23), safety is only scarcely mentioned in two of them. Out of the 21 typical activities of the Preliminary Design and Technology Completion Phase [5] (p.24), only one is related to safety. In the European Cooperation for Space Standardization (ECSS) brochures, the boundaries are not always clearly defined; for instance, product assurance includes reliability, availability, maintainability and safety activities [6] (p.15). Related to safety, an ECSS brochure is dedicated to this subject [12]. Safety is also scarcely introduced in IEEE Std. 1233 [13] (p.9-10). Textbooks related to systems engineering are many [11], [14], [15]. Of the few reviewed, all of them mention safety as an important requirement for the development of a complex system. For instance, Sage and Rouse consider safety as part of the "scientific and engineering effort" of SE, as transverse activities beside "reliability, maintainability, survivability, human engineering, and other factors" [11] (p.13). It is also worth mentioning that none of the resources cited above introduces the two equivalent concepts of ALARA (As Low As Reasonably Achievable) or ALARP (As Low As Reasonably Practicable).

In conclusion, it should be highlighted that systems engineering complements project management practices by providing means to communicate and coordinate the technical dimension of the project. However, while systems engineering is particularly well suited for the conception and development of complex products that are made of subprojects of a mechatronics nature (mechanics, electronics, controls software), it has not been designed for complex facilities.

## THE openSE FRAMEWORK

*What Is openSE?*

openSE is the systems engineering framework that CERN and a few particle physics research institutes have worked out to streamline project management processes. More specifically, it is intended to provide means to efficiently manage development projects of complex systems subject to or emitting ionizing radiation, while paying particular attention to four important aspects that may otherwise be omitted because of the natural focus that is naturally given by the project team to the project deliverable itself.

*openSE Governing Principles*

Five principles governed the development of openSE, namely: openness, leanness, participation, modularity and scalability [16] (pp. viii and ix).

Openness. The aim of research projects consists of creating knowledge and, in fundamental research at least, of disseminating and sharing it widely in a non-commercial framework.

Leanness. Because project and systems engineers and designers who are the primary group of beneficiaries of openSE are not management professionals but engineering experts, they are not necessarily keen to spend time on paperwork. Hence, it is necessary that the managerial tasks are kept to a minimum level, i.e. "lean".

Participation. The development and operation of scientific facilities and systems are usually performed in participative environments, i.e. all project and systems engineers and designers are expected to contribute actively to managerial tasks such as gathering requirements from customers or users, conducting risk analyses, planning and scheduling, reporting progress, etc.

Modularity. openSE is intended to cope with systems engineering requirements of a wide variety of projects. Projects may not need to implement all the features.

Scalability. Finally, openSE is equally scalable from large-scale scientific facility projects to equipment development or upgrade projects.

### The Core Components of openSE

As most of the management frameworks, openSE defines a lifecycle and its key processes, roles and responsibilities and key deliverables, mainly documents and decisions.

The openSE lifecycle is definitely inspired from various lifecycle models one can find in the literature (and especially the HERMES 4 project management methodology one [17]), but adapted to practices one can encounter in scientific facilities [18]. Contrary to a few proposed project lifecycles that are expressed in a rather complicated way, the decision was taken to rely on a rather simple model (see figure 1). The six phases of the openSE lifecycle are namely the Initialize, Study, Design, Build, Commission and Finalize phases. The scope and goals of each of these six phases are given bellow.

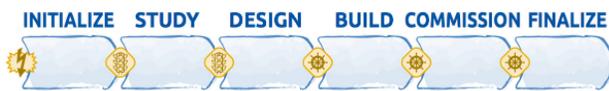

Figure 1: The openSE lifecycle.

The Initialize phase has three key goals:
- Analysing the present situation and defining what is the "problem" to solve;
- Proposing a few possible solution to the problem;
- Formalizing the decision to perform the project, at least to launch the project front-end phase.

The key deliverable of this phase is the Project Proposal which after being endorsed by a so-called Project Board may become the Project Roadmap.

The Study phase has four key goals:
- Gathering the needs, i.e. the users' requirements or the stakeholders' requirements;
- Converting the gathered needs into requirements;
- Identifying all possible solutions to the problem;
- Proposing one solution, i.e. the preferred solution, and demonstrating its feasibility by means of a Conceptual Design Report.

The Design phase has five key goals:
- Finalizing the definition of the needs;
- Finalizing the list of requirements accordingly;
- Designing the solution, i.e. performing the engineering design, also called the basic design or the systems-level design;
- Planning further the Build and Commission phases;
- If required, developing prototypes, mock-ups;

The key deliverable of this phase is the Technical Design Report.

The Build phase has three key goals:
- Performing the detailed design;
- *Materializing* the equipment, systems and, by the way, the facility; practically, this consists of procuring, manufacturing, assembling, installing, etc.;
- Verifying the conformity of the materialization, i.e. controlling that all the requirements have been correctly implemented.

The Commission phase has five key goals:
- Validating the outcome(s) of the project, i.e. demonstrating that all the users' requirements or the stakeholders' requirements are satisfied;
- Refining, i.e. getting rid of all the minor and not fully solved problems encountered during the previous phases, and ramping-up;
- If required, adapting the project to the evolving context;
- Training the operations and maintenance teams;
- Releasing Operations & Maintenance Documentation.

The Finalize phase has just one goal: capitalizing the lessons learned all along the project so that problems are not repeated again and again.

In matter of roles and responsibilities, two roles are critical and emphasised in openSE, the Project Board and the Project Manager. These roles definition are inspired from those defined in the HERMES project management methodology.

The responsibilities of the Project Board consist of:
- Ensuring the strategic management of the project;
- Guaranteeing the acquisition and availability of resources, in importance and in due time;
- Because of the latter, being ultimately responsible for the successful completion of the project;
- Validating the gates between phases, but also within phases when such gates are considered;
- And, in case of conflict, arbitrating.

The responsibilities of the Project Manager consist of:
- Ensuring the operational management of the project;
- Being responsible towards the project board for the organization of the project and its coordination.

## CONCLUSION

In this article, we have shown that it is possible to assemble the key concepts from major methodologies to define a project management and systems engineering framework suited to scientific projects. The attempts to define a dedicated framework is not new (see [19, 20] for instance), but this time and while the focus is placed on complex projects, openSE is designed in such a way that it is applicable to all types of projects of varying levels of complexity. Insights on openSE can be found at:

    http://cern.ch/openSE

## ACKNOWLEDGEMENT

The research leading to this framework has received funding from the European Commission under the FP7 ITN project PURESAFE, grant agreement no. 264336.

Proceedings of IPAC2016, Busan, Korea　　　　WEPMR055

# REFERENCES

[1] US Project Management Institute, http://www.pmi.org

[2] P. Bonnal, M. Baudin, and J.-M. Ruiz, "Systems Engineering and Safety Issues in Scientific Facilities Subject to Ionizing Radiations," *International Journal of Advanced Robotic Systems*, vol. 10, p. 10, 2013

[3] *A Guide to the Project Management Body of Knowledge*, 4$^{th}$ ed., Project Management Institute, Newton Square, PA, USA, 2008

[4] *Guidance on Project Management*, International Organization for Standards (ISO), Geneva, Switzerland, Standard ISO 21500:2012, 2012

[5] *NASA Systems Engineering Handbook*, NASA Headquarters, Washington, DC, 2007

[6] *Space Engineering: Policy and Principles*, European Coordination for Space Standardization (ECSS), Noordwijk, NL, Standard ECSS-E-00A, 1996

[7] International Council on Systems Engineering (INCOSE) http://www.incose.org

[8] *Guide to the Systems Engineering Body of Knowledge*, INCOSE, http://g2sebok.incose.org

[9] D. Milosevic and P. Patanakul, "Standardized project management may increase development projects success," *International Journal of Project Management*, vol. 23, no. 3, pp. 181–192, Apr. 2005

[10] D. Dvir, T. Raz, and A. J. Shenhar, "An empirical analysis of the relationship between project planning and project success," *International Journal of Project Management*, vol. 21, no. 2, pp. 89–95, Feb. 2003

[11] A. P. Sage and W. B. Rouse, Eds., *Handbook of Systems Engineering and Management*. Wiley-Interscience, 1999

[12] *Space Product Assurance: Safety*, European Coordination for Space Standardization (ECSS), Noordwijk, NL, Standard ECSS-Q-40A, 1996

[13] *Guide for Developing System Requirements Specifications*, The Institute of Electrical and Electronics Engineers, Inc. (IEEE), New-York, NY, Standard IEEE Std. 1233, 1998

[14] A. Kossiakoff and W. N. Sweet, *Systems Engineering: Principles and Practices*. Chichester: John Wiley & Sons, 2003

[15] R. Stevens, P. Brook, K. Jackson, and S. Arnold, *Systems engineering : coping with complexity*. London: Prentice Hall Europe, 1998

[16] *The openSE Framework*, PURESAFE ITN Project Community, CERN, Geneva, Switzerland and TUT, Tampere, Finland, 2014, http://cern.ch/openSE

[17] *HERMES Foundations: Management and Execution of Projects in Information and Communication Technologies*, Swiss Federal Strategy Unit for Information Technology, Bern, Switzerland 2003

[18] Ø. Husby, "Conceiving a Lean and Participative Project Management Framework Suited to Large-Scale Scientific Projects: How to Adapt Existing Systems Engineering and Project Management Best Practices to CERN's Projects?", M.Sc. thesis, Norwegian University of Science and Technology (NTNU), Trondheim, Norway, May 2013

[19] G. Bachy and A.-P. Hameri, "What to be implemented at the early stage of a large-scale project," International Journal of Project Management, vol. 15, no. 4, pp. 211–218, Aug. 1997

[20] G. Bachy, A.-P. Hameri, G. Lindecker, and M. Nordberg, "L' évolution de l'organisation dans un environnement multiprojet à grand échelle," in *Projectique: à la recherche du sens perdu*, Paris, France: Économica, 1996, pp. 185–196